\documentclass[useAMS,usenatbib]{mn2e}

\usepackage[T1]{fontenc}
\usepackage{aecompl}

\usepackage{perpage}  
\MakePerPage{footnote}  

\usepackage{mathptmx}  
\usepackage{threeparttable,threeparttablex,multirow,booktabs,longtable,rotating}  
\usepackage{lscape}  
\usepackage{afterpage}  
\usepackage{moresize} 
\usepackage{graphicx}   
\usepackage{amsmath,amssymb}
\usepackage{bm}  
\newcommand\numberthis{\addtocounter{equation}{1}\tag{\theequation}}
\usepackage{color}
\usepackage[ampersand]{easylist}
\usepackage[small,bf,singlelinecheck=false]{caption}  
\usepackage{float}  
\usepackage{placeins}  

\mathchardef\mhyphen="2D
\newcommand\numinsys{n_i}
\newcommand\multiplicitydistribution{N_{pl}}
\newcommand\trueplanets{\bm{N_k}}
\newcommand\detectedplanets{\bm{\hat{N_k}}}
\newcommand\detectedplanetsgrid{\bm{\hat{N}_{\bm{k,ij}}}}
\newcommand\transdurationratio{\bm{\xi}}
\newcommand\transdurationratiogrid{\bm{\xi_{\bm{ij}}}}
\newcommand\meanplanets{\mu}
\newcommand\invplane{\langle \theta \rangle}

\newcommand\teffmin{4100}
\newcommand\teffmax{6100}
\newcommand\rstarmax{1.15}
\newcommand\cdppmax{200}
\newcommand\baselinemin{1000}
\newcommand\dutycyclemin{0.6}
\newcommand\stellarcutnum{63,128}
\newcommand\samplecandidates{1,077}
\newcommand\samplesystems{816}
\newcommand\multiplicitysample{[631, 127, 45, 9, 3, 1]}
\newcommand\samplestellarcuts{1,790}
\newcommand\initialsample{4,150}
\newcommand\numcandidates{1,338}
\newcommand\numnotdispositioned{4,401}
\newcommand\numdispositionschanged{2,812}
\newcommand\periodmax{200}
\newcommand\periodmin{3}
\newcommand\mesmin{10}
\newcommand\radiusmin{1}
\newcommand\radiusmax{5}
\newcommand\nummodelsystems{$10^6$}

\newcommand\numplstep{0.1}
\newcommand\parameterstep{0.1}
\newcommand\numgridpoints{1,500}

\title[Flat Inner Disk Model as Alternative to Kepler Dichotomy]{A Flat Inner Disk Model as an Alternative to the Kepler Dichotomy in the Q1$-$Q16 Planet Population}

\author[T. Bovaird and C. H. Lineweaver]{T. Bovaird$^{1,2}$\thanks{E-mail: timothy.bovaird@anu.edu.au} and C. H. Lineweaver$^{1,2,3}$ \\ 
$^{1}$Research School of Astronomy and Astrophysics, Australian National University, Canberra, ACT 2611, Australia \\ 
$^{2}$Planetary Science Institute, Australian National University \\ 
$^{3}$Research School of Earth Sciences, Australian National University}

\begin{document} 
\pagerange{\pageref{firstpage}--\pageref{lastpage}} \pubyear{2016}

\maketitle 
\label{firstpage}

\begin{abstract}
We use simulated planetary systems to model the planet multiplicity of Kepler stars.
Previous studies have underproduced single planet systems and invoked the so-called Kepler dichotomy, 
where the planet forming ability of a Kepler star is dichotomous, producing either few or many transiting planets. 
In this paper we show that the Kepler dichotomy is only required when the inner part of planetary disks are just assumed to be flared.
When the inner part of planetary disks are flat, 
we reproduce the observed planet multiplicity of Kepler stars without the need to invoke a dichotomy.
We find that independent of the disk model assumed,			
the mean number of planets per star $\meanplanets\approx2$ for orbital periods between 3 and 200 days, 
and for planetary radii between 1 and 5 Earth radii. 	
This contrasts with the Solar System where no planets occupy the same parameter space.
\end{abstract}

\section{Introduction}
The Kepler Q1$-$Q16 catalog \citep{Mullally2015} uses 47 months of Kepler data collected from $\sim$190,000 stars. 
This has resulted in the detection of over 4,000 planet candidates orbiting $\sim$3,200 stars. 
While the majority of the $\sim$3,200 stars contain a single detected planet, 
transit signals from multiple planets have been detected around 656 of these stars.
Comparisons to the architecture of the Solar System are limited,
due to the relatively smaller periods and planetary radii that Kepler can efficiently sample.

\subsection{The Kepler Dichotomy}
\label{sec:dichotomy}

The mutual inclination distribution between planets around Kepler stars has been well studied
(\cite{Lissauer2011,Fang2012a,Figueira2012,Tremaine2012,Johansen2012,Weissbein2012,Ballard2014,Fabrycky2014}, see Appendix \ref{sec:prev_studies}). 
The majority of of these studies show good agreement between simulated planets and the Kepler sample when the orbital planes of simulated planets are closely aligned.
Specifically, when the mutual inclinations between planets are drawn from a Rayleigh distribution (a `flared' planetary disk) with a mode of the flare angle between $\sim1^\circ-5^\circ$. 

In contrast to the agreement for mutual inclinations, 
some studies report a significant underproduction of simulated systems with a single detected planet \citep{Lissauer2011,Johansen2012,Hansen2013,Ballard2014}. 
These studies underproduce the number of simulated stars with a single detected transiting planet by a factor of $\sim3$.

The underproduction of simulated systems with a single detected transiting planet 
has led to the proposal of dichotomous planetary systems in the Kepler field,
the so-called Kepler Dichotomy.
One population of planetary systems is required to either suppress planet formation, or be ``dynamically hot'' \citep{Hansen2013}, 
where mutual inclinations between planets are increased, or where planets are more likely to be ejected from the system. 
For the host stars in these planetary systems, the probability of detecting multiple transiting planets is reduced, 
leading to a higher proportion of stars with a single detected transiting planet in this population. 

Potential explanations for the dynamically hot planetary system population include dynamical instability caused by high mass planets \citep{Johansen2012,Lai2016}, 
instability or suppressed planet formation caused by stellar binaries \citep{Ballard2014},
varying surface density profiles and disk masses \citep{Moriarty2015},
varying strengths of gas depletion
or spin-orbit misalignment between the star and planet \citep{Spalding2016}.
\cite{Ballard2014} show that to account for the excess of detected single-planet transiting systems around M dwarfs,
these stars with a reduced probability of multiple transiting planets need to account for $\sim55$ per cent of M dwarfs in the Kepler field.

\begin{figure*} 
\includegraphics[width=0.95\textwidth]{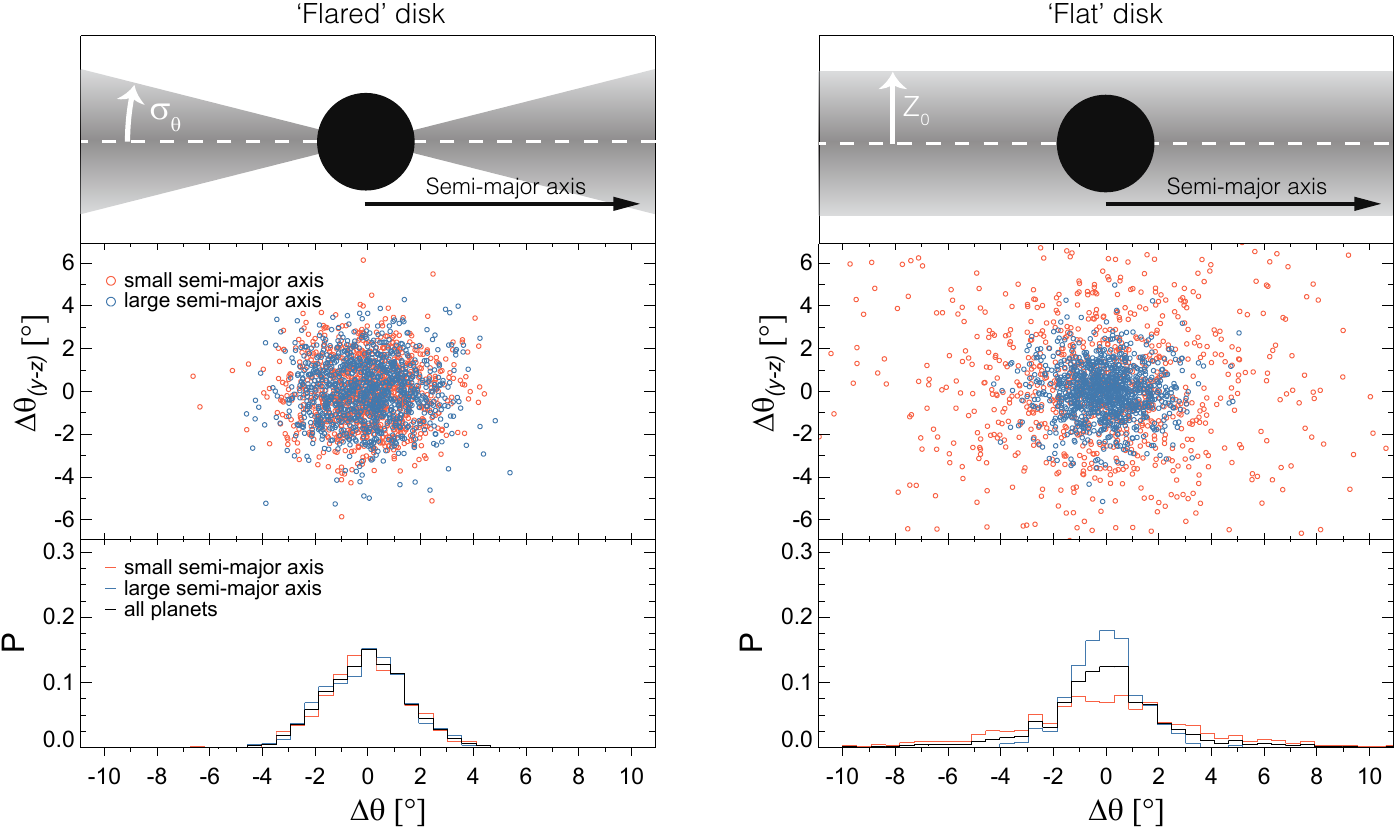} 
\caption{
Comparing the two disk models from Section \ref{sec:disk_model}, a `flared' (left) and `flat' (right) disk. 
The \emph{top} panels illustrate the two models viewed edge-on, i.e. perpendicular to the invariable plane (indicated by the white dashed line). 
The simulated inclinations (see Fig. \ref{fig:inclination_angles}) of planets with small ($0.05-0.3$ AU, red) and large ($0.3-0.6$ AU, blue) semi-major axes (around a Solar mass star) are shown  in the \emph{middle} panel for each model.
Under the flared disk model, the distribution of inclinations is independent of semi-major axis, 
whereas under the flat disk model, planets with a smaller semi-major axis tend to have a larger inclination and vice-versa.
The \emph{bottom} panel displays the distribution for the one component of inclination that the transit method is sensitive to (Section \ref{sec:mutual_inclinations} and Fig. \ref{fig:inclination_angles}).} 
\label{fig:model_illustration} 
\end{figure*}

\subsection{Detected Transiting Planets}

We define the true planetary system multiplicity vector $\trueplanets$ as the number of stars which are host to $k$ planets.
For the Kepler mission (and any transit survey), 
the observed vector $\detectedplanets$ will be significantly lower than the true $\trueplanets$, 
due to the low probability of planets transiting their host star, 
and since Kepler can only efficiently detect planets across a small fraction of parameter space. 
The observed planetary system multiplicity vector, hereafter simply referred to as the multiplicity vector, is given by
\begin{equation}
\detectedplanets = [\hat{N_1},\hat{N_2},\hat{N_3},..],
\label{eq:system_multiplicity}
\end{equation}
where $\hat{N_1}$, $\hat{N_2}$ and $\hat{N_3}$ are the the number of stars with 1, 2 and 3 \emph{detected} transiting planets respectively, and so on.
For the Kepler Q1$-$Q16 catalog (see Section 2.1), $1\le k \le 6$ and $\detectedplanets=[$2608, 413, 141, 52, 18, 3].

\subsection{Mutual Inclinations}
\label{sec:mutual_inclinations}

For two or more planets in the same planetary system, the mutual inclination between those planets is defined as the angle between their orbital planes.
The probability of multiple planets transiting the same star is non-negligible for small mutual inclinations only,
generally on the order of a few degrees.
Planets in the system with larger mutual inclinations, relative to the transiting planets, require alternative detection methods. 

In general, the true inclination of the orbital plane of a transiting planet cannot be determined from a transit lightcurve alone.
The transit method is only sensitive to the line-of-sight component of the inclination $i$ (Figure \ref{fig:inclination_angles}). 
In Figure \ref{fig:model_illustration}, the distribution of the detectable inclination component for a set of simulated planets is shown in the bottom panels. 
The orthogonal component of inclination, typically not detectable by the transit method, represents the y-axis of the middle panels. The true mutual inclination between a pair of planets is given by $\sqrt{\Delta\theta^2+\Delta\theta_\text{(y-z)}^2}$.

\subsection{An Alternative Disk Model}
\label{sec:disk_model}
In the studies mentioned in Section \ref{sec:dichotomy}, the true mutual inclinations between simulated planets are drawn from a Rayleigh distribution with mode $\sigma_{\Delta\phi}$. 
The Rayleigh distribution is composed of two Gaussian distributed components, with standard deviations equal to $\sigma_{\Delta\phi}$.
We can visualize the inclination distribution by considering one of these Gaussian components, i.e. viewing systems edge-on at an arbitrary plane perpendicular to the invariable plane, as in the top panels of Figure \ref{fig:model_illustration}. 
Rayleigh distributed mutual inclinations represent a `flared disk' model, where a planet's height above the invariable plane\footnote{
The mode of the Rayleigh distribution of inclinations relative to an invariable plane $\sigma_\phi$, 
is related to the Rayleigh distribution of mutual inclinations $\sigma_{\Delta\phi}$, by $\sigma_\phi \approx \sigma_{\Delta\phi}/\sqrt{2}$.
}
tends to increase with increasing semi-major axis. 
Planet inclinations relative to the invariable plane
do not depend on semi-major axis.

In this paper, we use a `flat disk' model,
where a planet's height above the invariable plane does not depend on semi-major axis,
and planet inclinations relative to the invariable plane tend to decrease with increasing semi-major axis
(as seen in the right-side panels of Figure \ref{fig:model_illustration}).
\cite{Hansen2013} tested the in situ assembly of close in planets, 
and found that planets with small semi-major axes tended to have larger inclinations, particularly $<0.1$ AU. 

\begin{figure} 
\includegraphics[width=0.47\textwidth]{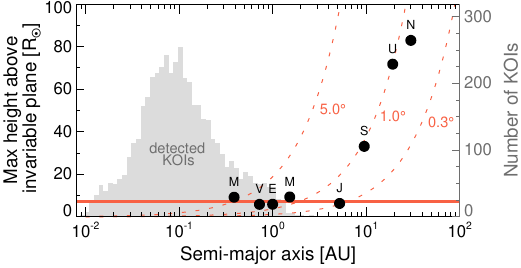} 
\caption{The maximum height above the Solar System's invariable plane over a single orbit for Solar System planets. 
Interior to Jupiter, the height above the invariable plane is approximately constant, represented by the red line with a value $Z_\text{max}\approx8 \;R_\odot$. 
Exterior to Jupiter, the height above the invariable plane flares out with increasing semi-major axis. 
Dashed lines of constant inclination relative to the invariable plane, are shown for 0.3, 1 and 5 degrees.
The semi-major axis distribution of Kepler Q1$-$Q16 planet candidates (gray histogram) falls well within the Solar System's constant-height regime.} 
\label{fig:solsys_heightaboveplane} 
\end{figure}

We apply this flat disk model to the typical semi-major axis space probed by Kepler, 
i.e. the interior part of planetary disks, as shown in Figure \ref{fig:solsys_heightaboveplane}.
In general, this represents planets with semi-major axes much less than the semi-major axes of the inner Solar System planets. 
We improve on previous modeling efforts by removing the flared disk assumption.
We show that for a flat inner planetary disk there is no need to invoke a dichotomous planetary system population, 
where one population of host stars have a decreased probability of hosting multiple transiting planets.

In Section \ref{sec:sample} we define our stellar and planetary samples, based on minimizing false positives and false negatives.
In Section \ref{sec:simulations} we estimate the transit and detection completeness across our parameter space.
In Section \ref{sec:period_radius_distributions} we estimate the underlying orbital period and planet radius distributions.
We outline the process of producing model planetary populations in Section \ref{sec:model_systems}.
In Section \ref{sec:compare},
we compare the simulated detections in our model systems to the Kepler Q1$-$Q16 candidates, for both flat and flared disk models.
In Section \ref{sec:results} we discuss the results from our model planet populations, 
including estimates for the mean number of planets per star within our parameter space.

\section{Sample selection}
\label{sec:sample}

When we generate a model planetary system, the stellar properties for that system are assigned from a random Kepler star in our stellar sample.
We produce our input stellar sample in the following way.

We begin with the 198,917 stars from the Kepler Q16 stellar catalog\footnote{
http://exoplanetarchive.ipac.caltech.edu/cgi-bin/TblSearch/nph-tblSearchInit?app=ExoTbls\&config=keplerstellar}. 
We limit our sample to low-noise Solar type stars, similar to the majority of previous studies mentioned in Section \ref{sec:dichotomy}.
We apply the following cuts to the input catalog:
\begin{align*}
\teffmin\; \text{K} <\; T_\text{eff}\; &< \teffmax\; \text{K} \\
\sigma_{\text{CDPP}_{45}} &< \cdppmax \text{ ppm} \\
R_* &< \rstarmax\; R_\odot \numberthis \label{eq:stellar_sample} \\
T_\text{baseline} &> \baselinemin \text{ days} \\
f_\text{duty} &> \dutycyclemin 
\end{align*}
where $T_\text{eff}$ and $R_*$ are the stellar effective temperature and radius respectively. 
The 4.5 hour CDPP (combined differential photometric precision, \cite{Christiansen2012}) of the star, 
a measure of the combined instrumental and stellar noise, is given by $\sigma_{\text{CDPP}_{45}}$.
$T_\text{baseline}$ is the timespan of observations for each star and $f_\text{duty}$ is the fraction of valid observations over $T_\text{baseline}$.
Note that the combination of $T_\text{baseline}$ > 1000 days and $f_\text{duty}$ > 0.6 generally ensures at least 3 transits for orbital periods up to 200 days. 

The above stellar cuts, in addition to removing stars without a reported mass, results in our input stellar sample of \stellarcutnum\ stars.
In later sections, the stellar properties of each simulated planetary system are assigned from a randomly drawn star in this sample.

To minimize the detection incompleteness and false-positives in our observed planet sample, to which we will compare our simulations, we select only those planets with a high pipeline detection efficiency.
We begin with the \numcandidates\ Kepler Objects of Interest (KOIs) labeled `candidate' by the Q1$-$Q16 pipeline\footnote{
http://exoplanetarchive.ipac.caltech.edu/cgi-bin/TblView/nph-tblView?app=ExoTbls\&config=koi}.
An additional \numnotdispositioned\ KOIs are labeled `not dispositoned'. We update these dispositions using the Kepler Q17 catalog for reference.
This results in \numdispositionschanged\ KOIs changing from `not dispositioned' to `candidate'.

These planets form our initial sample of \initialsample\ planet candidates from the Q1$-$Q16 catalog \citep{Mullally2015}.
To conform with our input stellar sample, we remove planets around host stars outside of our stellar parameter space defined by Equation \ref{eq:stellar_sample}.
This reduces our sample of observed planets from \initialsample\ to \samplestellarcuts. 

We set an upper orbital period limit of \periodmax\ days to avoid the increase in false-positives towards the Kepler orbital period of $\sim\,$372 days \citep{Mullally2015}, 
and to remain consistent with the Kepler pipeline completeness calculations \citep{Christiansen2015}.

The Kepler pipeline is known to have an increasing false-negative rate with decreasing orbital period for orbital periods $\lesssim 3$ days.
This is largely due to the pipeline harmonic filter, which can remove transit signals which are on the same timeframe as the expected stellar noise \citep{Christiansen2015}.
In addition, a small fraction of the fitted planetary radii for planets with orbital periods $\lesssim$ 10 days can be significantly lower than the true planet radius, diluting the transit signals for some of these planets.
We choose a orbital period lower limit of \periodmin\ days, in order to retain a sample of Kepler stars with $\ge 4$ transiting planets.

The Kepler pipeline reports a summary statistic for the strength of a transit detection, the Multiple Event Statistic (MES). 
A lower limit planet radius of $\radiusmin\;R_\oplus$ and a lower limit MES of \mesmin\ are chosen since false-positives are dominated by low MES ($\lesssim 8$) detections \citep{Mullally2015}. 
An upper planet radius limit of $\radiusmax\;R_\oplus$ is chosen to avoid increasing false-positives with planet size, and since the mass-radius relation becomes degenerate for larger planetary radii. 

To summarize, we only retain the Kepler Q1$-$Q16 candidates which meet the following criteria:
\begin{align*}
\periodmin\; \text{days} <\; &P < \periodmax\; \text{days}\\
\radiusmin\; \text{R}_\oplus <\; &R_p < \radiusmax\; \text{R}_\oplus \numberthis \label{eq:planet_sample} \\
M&ES > \mesmin
\end{align*}
This results in our observed sample of \samplecandidates\ candidates in \samplesystems\ planetary systems, within the parameter space outlined in Equations \ref{eq:stellar_sample} and \ref{eq:planet_sample}.
The observed planetary system multiplicity vector $\detectedplanets$ (Equation \ref{eq:system_multiplicity}) for our parameter space is given by
\begin{equation}
\label{eq:multiplicity_observed}
\detectedplanets = \multiplicitysample.
\end{equation}

\section{Transit and detection efficiency}
\label{sec:simulations}

When attempting to estimate the underlying multiplicity vector $\trueplanets$ given the observed $\detectedplanets$ (Equation \ref{eq:system_multiplicity}), 
there exists a degeneracy between the underlying multiplicity distribution and the underlying mutual inclination distribution.
For example, an observed $\detectedplanets$ could be reproduced by systems which contain many planets with a large dispersion in mutual inclinations, 
or by systems containing fewer planets but with a small mutual inclination dispersion. 
These two underlying distributions must be modeled simultaneously.

We estimate the underlying inclination and multiplicity distributions of Kepler systems in the Q1$-$Q16 catalog, 
within the period and radius parameter space where Kepler can more reliably detect transiting planets 
(given by Equations $\ref{eq:stellar_sample}$ and $\ref{eq:planet_sample}$).
We produce sets of \nummodelsystems\ simulated planetary systems across a grid of inclination and multiplicity distributions.
For each set of model assumptions, we estimate the probability of Kepler detecting each simulated planet. 
By comparing the $\detectedplanets$ for a set of simulated systems to the $\detectedplanets$ for the Q1$-$Q16 Kepler catalog, 
we can estimate the underlying architecture between planetary orbital planes and the distribution of the number of planets per star.

\subsection{Pipeline detection efficiency}
\label{sec:pipeline_completeness}

\begin{figure} 
\includegraphics[width=0.47\textwidth]{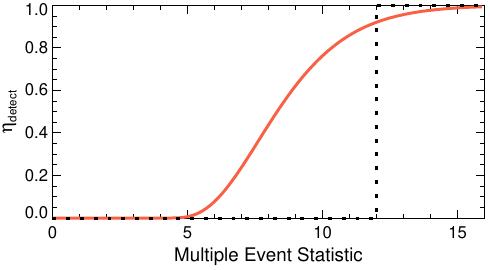} 
\caption{The Kepler Q1$-$Q16 pipeline detection completeness as a function of MES (Multiple Event Statistic, Equation \ref{eq:MES}), from \protect\cite{Christiansen2015}. 
The black dashed line represents a typical completeness step function assumed in previous studies.
The full form of the pipeline detection completeness can be seen in Equation \ref{eq:eta_disc}.} 
\label{fig:pipeline_completeness} 
\end{figure}

Detection incompleteness and false-positives are important issues when comparing the detected planets around simulated and observed stars. 
Previous studies did not have the benefit of the Kepler pipeline detection completeness provided by transit injection and recovery experiments \citep{Christiansen2015}, shown in Figure \ref{fig:pipeline_completeness}.

We follow the approximation of the pipeline MES by \citep{Burke2015}, which includes a limb-darkening approximation and accounts for nonzero impact parameters.
\begin{equation}
MES = \frac{0.84\,\delta(c+s\sqrt{\delta})}{\sigma_\text{cdpp}}\sqrt{n_\text{tr}}
\label{eq:MES}
\end{equation}
where $c=1.0874$ and $s=1.0187$ for G dwarfs, 
$\delta= (R_p/R_*)^2$, and $n_\text{tr}$ is the number of transits.
Values of $\sigma_\text{cdpp}$ are reported for 14 different transit durations from 1.5 hours to 15 hours for each Kepler star \citep{Burke2015}.
The $\sigma_\text{CDPP}$ value chosen for Equation \ref{eq:MES} is interpolated from the 14 reported CDPP values, 
to match the transit duration of the planet.

The number of transits for a planet is estimated by $n_\text{tr}=(T_\text{baseline}\times f_\text{duty}) / P$, where $P$ is the planet period,
$T_\text{baseline}$ is the total observing time for the Q1$-$Q16 catalog ($\sim\,$1426 days) 
and $f_\text{duty}$ is the duty cycle for the observed star; the fraction of valid observations over the observing baseline.
Note that $T_\text{baseline}$ and $f_\text{duty}$ are reported for each star, accounting for systematics such as the differences in CCD detectors and pixels.

We define $\eta_\text{detect}$ as the Kepler pipeline completeness, shown in Figure \ref{fig:pipeline_completeness}.
The pipeline completeness as a function of the multiple event statistic is approximately represented by the $\Gamma$ cumulative distribution function
\begin{equation}
\eta_\text{detect}(MES) = \frac{1}{c^b \Gamma(b)}\int_0^{MES-\beta} x^{b-1}e^{-x/c}dx
\label{eq:eta_disc}
\end{equation}
where $\Gamma$ is the Gamma function.
For our sample of FGK dwarfs, $b=4.35$, $c=1.05$ and $\beta=4.093$ \citep{Christiansen2015}.

To calculate the total probability of transit detection $\eta(P,R_p)$, 
we must also take into account the geometric transit probability $\eta_\text{transit}$ of a planet,
\begin{equation}
\eta_\text{transit} = \frac{R_*}{a}, 
\label{eq:eta_transit}
\end{equation}
where $a$ is the semi-major axis of the planet.
The product of these two equations gives the total transit detection probability $\eta(P,R_p)$,
the probability of the planet transiting $\eta_\text{transit}$,
and the probability of the transiting planet being detected by the Kepler Q1$-$Q16 pipeline $\eta_\text{detect}$,
\begin{equation}
\eta(P,R_p) = \eta_\text{transit} \times \eta_\text{detect}
\label{eq:eta_total}
\end{equation}

\begin{figure} 
\includegraphics[width=0.47\textwidth]{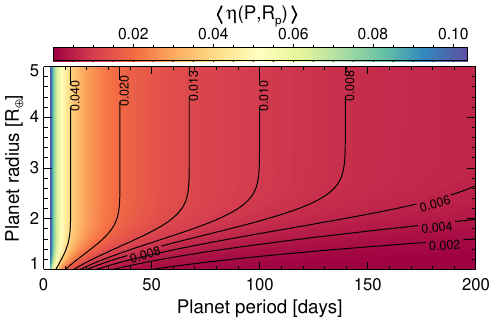} 
\caption{The probability of a planet transiting and being detected by the Kepler Q1$-$Q16 pipeline, $\langle\eta(P,R_p)\rangle$, 
over the planet period and planet radius parameter space of our simulations. 
The probability at each grid point is the average value across our stellar sample, the 63,128 stars which adhere to Equation \ref{eq:stellar_sample}.
For planets with radii exceeding $2.5R_\oplus$, the pipeline detection efficiency $\eta_\text{detect}\approx1$,
and $\langle\eta(P,R_p)\rangle$ is well estimated by the transit probability $\eta_\text{transit}$.
}
\label{fig:completeness_grid} 
\end{figure}

Given an input star from our stellar sample (Equation \ref{eq:stellar_sample}) and using Equations \ref{eq:MES} to \ref{eq:eta_total}, 
we can estimate the total transit and detection completeness of a simulated planet
with planet parameters of period, radius and transit duration.
Across our planetary parameter space used in this paper (Equation \ref{eq:planet_sample}),
we estimate the mean total transit and detection completeness $\langle\eta(P,R_p)\rangle$,
by taking the mean value of $\eta(P,R_p)$ at each grid point over all stars in our stellar sample.
This is shown in Figure \ref{fig:completeness_grid},
where $\langle\eta(P,R_p)\rangle$ ranges from $\sim0$ to a maximum of $\sim0.1$.
Transit and detection probabilities $>5$ per cent only exist for planets with orbital periods $\lesssim8$ days.
It can be seen that the pipeline detection probability becomes important for planetary radii less than $2.5$ Earth radii.

\section{Underlying planet distributions}
\label{sec:period_radius_distributions}

Our simulated results of the number of stars with $k$ detected transiting planets is reliant on an input planet radius and orbital period distribution, 
which has been one of the primary goals of the Kepler mission. 
The planet radius distribution is often modeled as a broken power law \citep{Youdin2011,Howard2012,Burke2015}, 
with a logarithmic plateau at $\lesssim 2.5R_\oplus$.
This logarithmic plateau is also seen when the pipeline efficiency is probed using transit injection and recovery experiments \citep{Petigura2013}.

For orbital periods between 50 and 300 days, a single power law is sufficient to describe the orbital period distribution \citep{Burke2015}. 
Our parameter space includes planets with orbital periods less than 50 days, 
where the transit and detection completeness is more dynamic,
particularly for periods $\lesssim15$ days (Figure \ref{fig:completeness_grid}).
For this parameter space, a single power law is not sufficient, 
and we model the orbital period distribution as a broken power law. 

The planet radius and orbital period distributions are combined into a planet distribution function (PLDF),
in this case composed of a broken power law for the distribution of orbital periods, 
and a broken power law for the distribution of planetary radii. 

Our PLDF has 7 free parameters, $F_0$, $\beta_1$, $\beta_2$, $P_\text{brk}$, $\alpha_1$, $\alpha_2$, $R_\text{brk}$, 
where $F_0$ is the number of planets per star within our parameter space, 
and $R_\text{brk}$ and $P_\text{brk}$ are the transition points between the two power laws for the planet radius and orbital period respectively, 
\begin{align*}	
&\frac{df}{dP\,dR_p} = C\,F_0\, g(P,R_p) \numberthis\label{eq:pldfmodel} \\ =
&\begin{cases}
C\,F_0\,P^{\beta_1} R_P^{\alpha_1} & \text{if }P < P_\text{brk}\text{ and }R_p < R_\text{brk}  \\ 
C\,F_0\,P^{\beta_1} R_P^{\alpha_2} R_\text{brk}^{\alpha_1-\alpha_2} & \text{if }P < P_\text{brk}\text{ and }R_p \ge R_\text{brk} \\
C\,F_0\,P^{\beta_2} P_\text{brk}^{\beta_1-\beta_2} R_P^{\alpha_1} & \text{if }P \ge P_\text{brk}\text{ and }R_p < R_\text{brk} \\
C\,F_0\,P^{\beta_2} P_\text{brk}^{\beta_1-\beta_2} R_P^{\alpha_2} R_\text{brk}^{\alpha_1-\alpha_2} & \text{if }P \ge P_\text{brk}\text{ and }R_p \ge R_\text{brk}
\end{cases}  
\end{align*}
where the power law exponents $\alpha_1$, $\alpha_2$ and $\beta_1$, $\beta_2$ represent the exponents for the orbital period and the planet radius distribution respectively, either side of the power law breaks. 

\begin{figure} 
\includegraphics[width=0.47\textwidth]{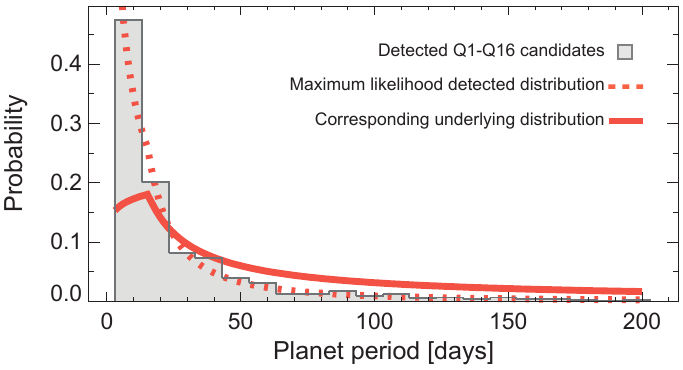} 
\caption{
The estimated underlying period distribution used for populating simulated planetary systems in this paper.
The underlying distribution is the marginalized orbital period PLDF maximum likelihood model (Equation \ref{eq:pldfmodel} and Table \ref{tab:underlying_distributions}). 
The expected detected distribution (dashed red line), 
results when the underlying distribution is convolved with the average total transit and detection probability $\langle\eta(P,R_p)\rangle$ for the stars in our sample.
This can be compared with the detected Q1$-$Q16 Kepler candidates (grey histogram).
}
\label{fig:debiased_period} 
\end{figure}

\begin{figure} 
\includegraphics[width=0.47\textwidth]{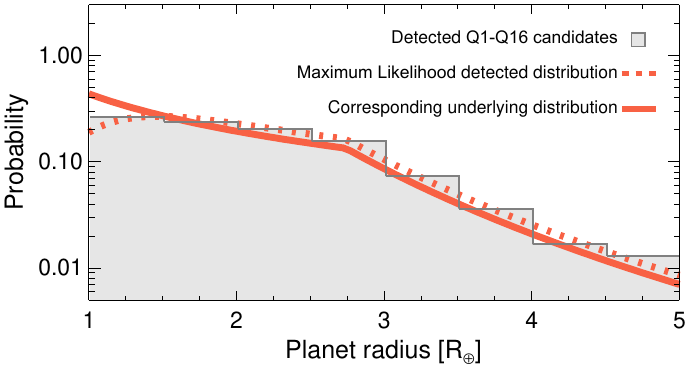} 
\caption{
The same as Figure \ref{fig:debiased_period}, except for the planet radius distribution.
The estimated underlying distribution is given by a broken power law with a break at $\sim$2.7 Earth radii (Table \ref{tab:underlying_distributions}).
}
\label{fig:debiased_radius} 
\end{figure}

For each set of model parameters in an underlying planet distribution function, 
an expected number of planet detections is computed by convolving the planet distribution function with $\langle\eta(P,R_p)\rangle$ 
(Figure \ref{fig:completeness_grid}).
The number of expected detections for an underlying planet distribution function is then compared to the number of Kepler Q1$-$Q16 detections
by maximizing the Poisson likelihood of the PLDF.
The maximum likelihood derivation for our PLDF (Equation \ref{eq:pldfmodel}) is shown in Appendix \ref{app:pldf}, 
with maximum likelihood parameters of
\begin{threeparttable}
\small
\caption{Maximum likelihood PLDF parameters} \label{tab:underlying_distributions}
\begin{tabular}{ccccccc}
\midrule
$F_0$ & $\beta_1$ & $\beta_2$ & $P_\text{brk}$ & $\alpha_1$ & $\alpha_2$ & $R_\text{brk}$ \\
\midrule
0.852 & 1.007 & -0.932 & 15.332 & -1.168 & -4.906 & 2.740 \\
\hline \\
\end{tabular}
\end{threeparttable}
indicating breaks in the power law distributions at $\sim$15 days and $\sim$2.7 $R_\oplus$ for orbital periods and planetary radii respectively.
The break in the orbital period distribution corresponds to a peak in the distribution,
whereas the break in the planet radius distribution corresponds to the logarithmic planet plateau for $R_p\lesssim2.7$ $R_\oplus$.
These results appear to be consistent with \cite{Foreman-Mackey2014}, 
where breaks in the logarithmic orbital period and planet radius rates are indicated at $\sim20$ days and $\sim2-3$ Earth radii respectively.

We can marginalize our maximum likelihood PLDF in terms of orbital period and planet radius.
This is shown in Figures \ref{fig:debiased_period} and \ref{fig:debiased_radius} respectively,
where the thick red lines represent our marginalized maximum likelihood PLDF, 
which is the estimated underlying planet distribution. 
The dashed red lines indicate the corresponding transit detected distribution, 
after applying the mean total transit and detection probability $\langle\eta(P,R_p)\rangle$ for the stars in our sample.

When producing model planetary systems in Section \ref{sec:model_systems},
we assign orbital periods and planetary radii by drawing randomly from the maximum likelihood underlying distributions,
shown in Figures \ref{fig:debiased_period} and \ref{fig:debiased_radius}.

\subsection{Parameterizing Planet multiplicity}
\label{sec:planet_multiplicity}

For a set of model planetary systems, we need to assume a distribution for the inherent number of planets per star within our parameter space, $\multiplicitydistribution$. 
In this paper we trial two different $\multiplicitydistribution$ distributions. 

The first trial distribution is a modified Poisson distribution $N_{pl,\text{Poi}}$ \citep{Fang2012a}. 
Each star is assigned a random number of planets, drawn from a Poisson distribution with mean $\meanplanets$. 
Stars drawn with zero planets are redrawn from the same Poisson distribution, 
until all \nummodelsystems\ model planetary systems are populated with planets,
resulting in a final mean $\ge\meanplanets$. 

The second trial distribution is a modified Exponential distribution $N_{pl,\text{exp}}$ \citep{Gaidos2016} and is produced in the same way as $N_{pl,\text{Poi}}$, 
except that stars are assigned a number of planets drawn from an exponential distribution with mean $\meanplanets$.
The mode of the exponential distribution is always 0, 
resulting in a natural tendency for more planetary systems to contain a single transiting planet rather than multiple transiting planets.
It has been shown that when an exponential distribution is used to model the inherent number of planets per star, 
no Kepler dichotomy is required \citep{Gaidos2016}.
We include this trial distribution for comparative purposes.

\section{Producing model planetary systems}
\label{sec:model_systems}
In our simulations, we assume two different planetary disk models, 
and two different distributions for the number of planets per star $\multiplicitydistribution$, 
resulting in simulations with four unique combinations of model assumptions. 
For a given set of model assumptions, we generate populations of planetary systems across a grid.
The mean of the number of planets per star $\meanplanets$ ranges from 0.5 to 3.5 in steps of \numplstep.
For the flared disk model, the mode of the Rayleigh distributed mutual inclinations ranges from 0 to 5 degrees in steps of \parameterstep\ degrees.
Similarly for the flat disk model, the standard deviation of the height above the invariable plane ranges from 0 to 5 $R_*$ in steps of \parameterstep\ $R_*$.

This results in a total of \numgridpoints\ grid points for each set of model assumptions, 
with \nummodelsystems\ model planetary systems generated at each grid point.
Each model planetary system is produced as follows:\\

\begin{easylist}
\ListProperties(Space*=0.1cm)
& A random star is chosen from our sample of \stellarcutnum\ Kepler stars outlined in Section \ref{sec:sample}, and its mass and radius are assigned to the star in the model system.

& The angle to the system's invariable plane relative to the observer $\invplane$ (Figure \ref{fig:inclination_angles}),  is chosen from a random point on a sphere: $0 \le \cos\left[\invplane\right] \le 1$.

& The number of planets in the system $\numinsys$ is drawn randomly, according to the assumed $\multiplicitydistribution$ distribution from Section \ref{sec:planet_multiplicity}, with a mean value based on the current grid point.

& The radii of the $\numinsys$ planets are drawn from the underlying distribution in Section \ref{sec:period_radius_distributions} ($\radiusmin\;\text{R}_\oplus < R_p < \radiusmax\;\text{R}_\oplus$), and converted to their corresponding masses\footnote{$M_p\approx\left(R_p/a\right)^b$, where $a$$\sim$1.11 and $b$$\sim$2.41.}.

& The periods of the $\numinsys$ planets are drawn randomly from the underlying distribution in Section \ref{sec:period_radius_distributions} ($\periodmin\;\text{days}< P< \periodmax\;\text{days}$), and converted to their corresponding semi-major axes, using the stellar properties of the assigned Kepler star.

& The dynamical stability of the system is estimated by testing the stability of sets of 3 sequential planets, or pairwise if $\numinsys=2$. 
If any set of planets in the system is deemed unstable, the system is labeled unstable and new planet periods for all $\numinsys$ planets are redrawn as in step 5.
See Section \ref{sec:stability} for a complete description of estimating the stability of a system, including termination criteria. 

& The inclinations of the planetary orbital planes relative to the observer, 
are determined from $\invplane$ (step 2) and the assumed planetary disk model (Section \ref{sec:disk_model}),
and the parameter value at the current grid point.
See Appendix \ref{sec:assign_inclinations} for a complete description of how planet inclinations are assigned for flat and flared disk models.
\end{easylist}

The above steps generate the \nummodelsystems\ model systems according to the assumed disk model, 
the assumed planet multiplicity distribution, 
the current grid point parameters, 
and the underlying planet period and radius distributions.
The final step is to estimate which simulated planets would be detected by the Kepler Q1$-$Q16 pipeline, 
and compare this detected sample to the observed Q1$-$Q16 detections.

\subsection{Determining transiting and detected planets}
\label{sec:detemine_detections}

Once planetary inclinations are assigned, the model system is complete and we test for transiting planets. 
We define a transiting planet by its impact parameter $b$, where a planet is defined to transit if:
\begin{equation}
b =\frac{a}{R_*}\cos i \le 1
\end{equation}
where $R_*$, $a$ and $i$ were determined from steps 1, 5 and 7 respectively.

For each simulated transiting planet, we estimate the Multiple Event Statistic ($MES$, Equation \ref{eq:MES}).
The $MES$ is dependent on stellar properties, along with the planet's orbital period, radius and transit duration.
Circular orbits are assumed when estimating transit durations. 
The planet's $MES$ is then used to estimate the pipeline detection efficiency $\eta_\text{detect}$ (Equation \ref{eq:eta_disc}).
For each simulated transiting planet, a uniform random number $Y_m$ is drawn between $0$ and $1$.
A simulated planet is labeled as detected if it transits, 
and if its pipeline detection efficiency $\eta_\text{detect} > Y_m$.
All simulated planets which meet this criteria are added to the detected planet sample for the grid point, ${\bm{X_{\bm{ij}}}}$, 
where $i$ and $j$ represent the current grid point.

\section{comparing Simulated and observed planet detections}
\label{sec:compare}

The simulations outlined in Section \ref{sec:model_systems} were performed across a grid for the 4 sets of model assumptions. 
For each grid point, the simulated planet detections ${\bm{X_{\bm{ij}}}}$ are used to generate two distributions,
the system multiplicity vector $\detectedplanetsgrid$ (Equation \ref{eq:system_multiplicity}),
and the distribution of orbit normalized transit duration ratios $\transdurationratiogrid$ \citep{Steffen2010}.
Unlike the $\detectedplanetsgrid$ distribution, 
the $\transdurationratiogrid$ distribution only consists of model systems with two or more detected transiting planets.
For a pair of planets orbiting the same star,
\begin{equation}
\label{eq:xi_main}
\xi=\frac{T_\text{dur,in}/P_\text{in}^{1/3}}{T_\text{out,in}/P_\text{out}^{1/3}},
\end{equation}
where $T_\text{dur}$ and $P$ are the transit durations and the periods for the inner and outer planets, 
given by the subscripts in and out respectively. 
For each unique planet pair in a system, $\xi$ is calculated, 
giving $\numinsys(\numinsys+1)/2$ values of $\xi$ for a star with $\numinsys$ planets.
For each grid point we generate the ensemble $\transdurationratio_{ij}$ distribution
by calculating the $\xi$ value for each unique pair of simulated transit detections orbiting the same star, 
across all \nummodelsystems\ model systems. 
For a deeper discussion of $\xi$, see Appendix \ref{app:xi}.

The $\detectedplanetsgrid$ and $\transdurationratiogrid$ distributions are compared to the $\detectedplanets$ and $\transdurationratio$ distributions of the Kepler Q1$-$Q16 candidates, 
and are used to assess the goodness of fit at each grid point.
We perform a $\chi^2$ goodness of fit test (Equation \ref{eq:chisqr}) 
comparing the simulated $\detectedplanetsgrid$ to the observed $\detectedplanets$ for our parameter space (Equation  \ref{eq:multiplicity_observed}).
We scale $\detectedplanetsgrid$ such that $\sum\detectedplanetsgrid=\sum \detectedplanets$.
To compensate for the poor quality of the $\chi^2$ test with low cell counts, values less than 5 are merged into their adjacent cells.
\begin{equation}
\label{eq:chisqr}
\chi^2 = \sum^n_{k=1} \frac{(\detectedplanets - \detectedplanetsgrid)^2}{\detectedplanetsgrid}.
\end{equation}

Similarly, we perform a two-sample Kolmogorov$-$Smirnov (KS) test between the simulated $\transdurationratio_{ij}$ distribution at each grid point, and the $\transdurationratio$ distribution of the observed Q1$-$Q16 Kepler candidates within our parameter space. 

\subsection{Flared Disk and Poisson distributed planets per star}

\begin{figure} 
\includegraphics[width=0.46\textwidth]{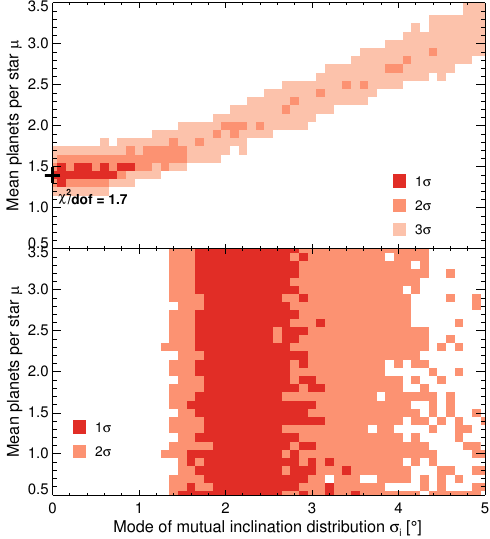} 
\caption{Comparing the simulated planet detections to the observed Q1$-$Q16 Kepler candidates, 
for Poisson distributed number of planets per star $\multiplicitydistribution$ with mean $\meanplanets$,
and a flared disk model (Rayleigh distributed mutual inclinations with mode $\sigma_i$).
\emph{Top:} 
Comparing the Kepler Q1$-$Q16 $\detectedplanets$ distribution (Equation \ref{eq:multiplicity_observed}) 
to the simulated $\detectedplanetsgrid$ distribution for each grid point.
Highly coplanar systems are favored. 
\emph{Bottom:} 
Comparing the Kepler Q1$-$Q16 $\transdurationratio$ distribution (Equation \ref{eq:xi_main}) to the simulated $\transdurationratiogrid$ distribution for each grid point. 
Contrary to comparing $\detectedplanets$ distributions, coplanar systems with $\sigma_i \lesssim 1.3^\circ$ are ruled out.
}
\label{fig:multiplicity_ks_grids_sigmai} 
\end{figure}

In the top panel of Figure \ref{fig:multiplicity_ks_grids_sigmai}, $\detectedplanetsgrid$ is compared to $\detectedplanets$ at each grid point,
under the assumption of a flared planetary disk and a Poisson distributed number of planets per star.
The $\chi^2$ values are represented by the 1$\sigma$, 2$\sigma$ and 3$\sigma$ values relative to the best fit.
As expected, no good fit is found to the $\detectedplanets$ distribution, 
as is the case in the majority of previous studies
\citep{Lissauer2011,Johansen2012,Ballard2014,Gaidos2016}.

The bottom panel displays the resulting p-values from the KS test between the $\transdurationratio_{ij}$ and $\transdurationratio$ distributions.
The orbital normalized transit duration ratios favor mutual inclinations with a mode between $1.5-4$ degrees,
consistent with all previous studies shown in Appendix \ref{sec:prev_studies}. 
The mean number of planet per star cannot be determined from the $\transdurationratiogrid$ distribution alone.

It is clear from Figure \ref{fig:multiplicity_ks_grids_sigmai} that the best-fit regions (dark red) of the two tests do not appear consistent. 
Comparing multiplicity vectors favors near coplanar mutual inclinations, with a mode $\lesssim1^\circ$ (top panel).
However, modes $\lesssim 1.5^\circ$ are ruled out by comparing orbit normalized transit duration ratios (bottom panel).

\begin{figure} 
\includegraphics[width=0.46\textwidth]{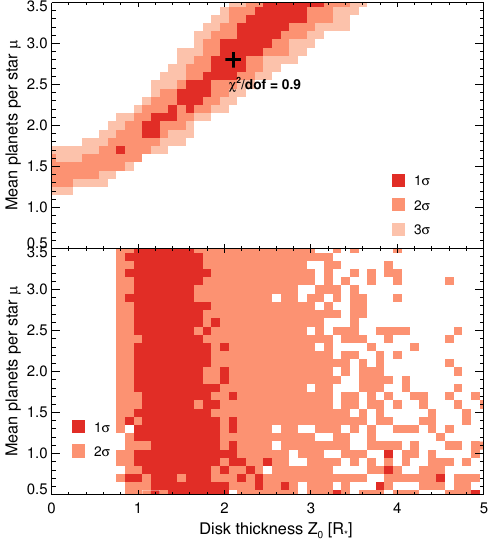} 
\caption{The same as Figure 7, except for a flat disk model instead of a flared disk model.
Both simulated $\detectedplanets$ and $\transdurationratio$ distributions are consistent with the detected Kepler Q1$-$Q16 candidates for
a flat disk model with Gaussian disk thicknesses between $\sim$1 and $\sim$3 stellar radii.
}
\label{fig:multiplicity_ks_grids_z0} 
\end{figure}

\subsection{Flat Disk and Poisson distributed planets per star}
For the set of simulations with a Poisson distributed $N_{pl}$ and a flat disk model,
the two tests appear more consistent (Figure \ref{fig:multiplicity_ks_grids_z0}).
Unlike for the flared disk model, the best fit $\detectedplanetsgrid$ is a good match to $\detectedplanets$,
giving a $\chi^2/\text{dof}$ of 0.9, indicating that no Kepler dichotomy is required.

While Gaussian disk thicknesses $>2\;R_*$ are supported by comparing multiplicity vectors,
comparing orbit normalized transit durations refines the disk thickness to $1\;R_*\lesssim Z_0 \lesssim 2\;R_*$. 
There is significant overlap between the two tests within this region. 

\begin{figure*} 
\includegraphics[width=0.95\textwidth]{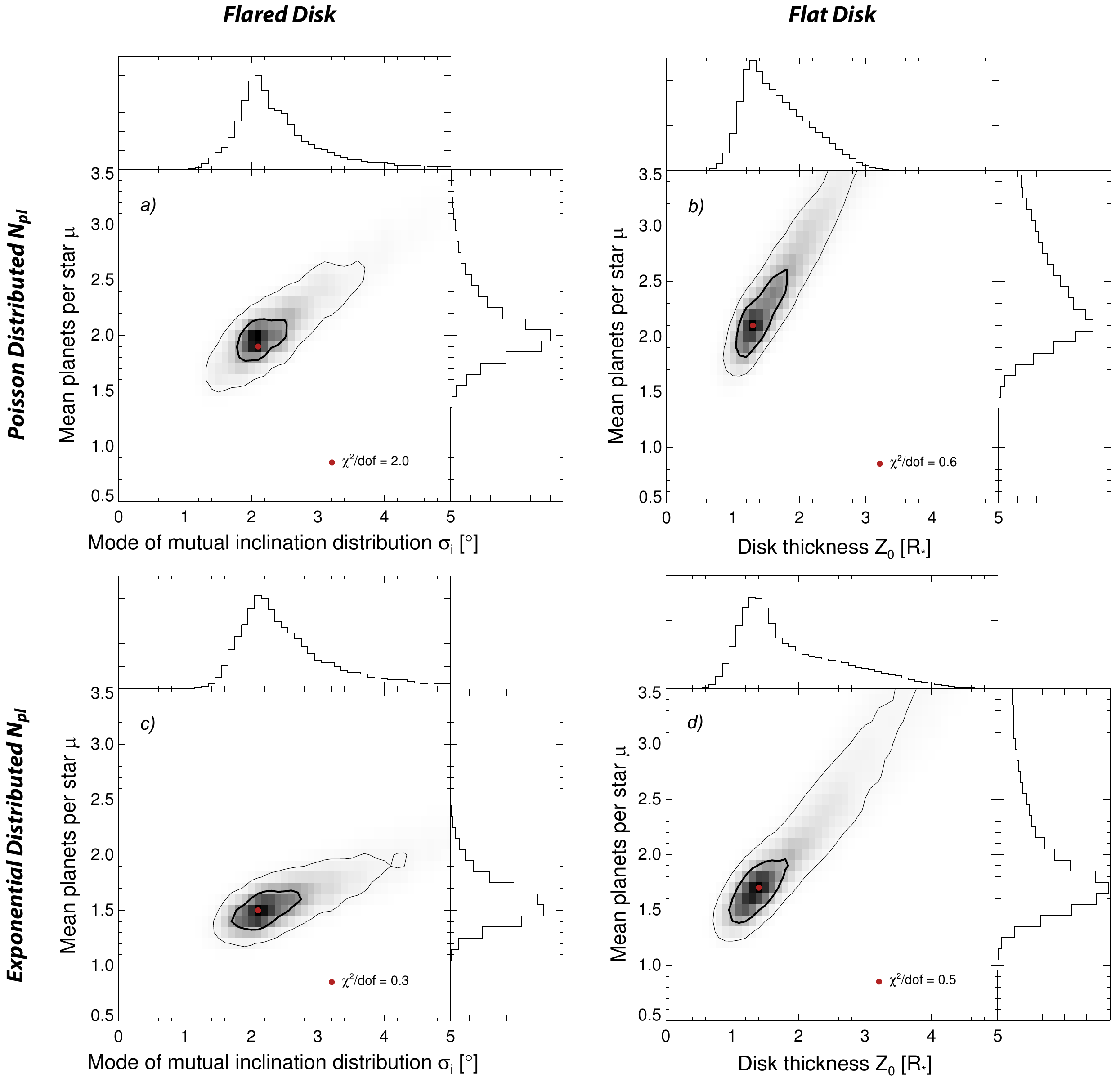} 
\caption{The combined probability distributions $P_\text{combined}$ for our 4 sets of model assumptions. 
$P_\text{combined}$ takes into account the goodness of fit between the simulated $\detectedplanetsgrid$ distributions
and the Kepler Q1$-$Q16 $\detectedplanets$, 
and the consistency of the simulated $\transdurationratiogrid$ distributions and Q1$-$Q16 $\transdurationratio$ distribution. 
For example, panel a) corresponds to the combined top and bottom panels of Figure \ref{fig:multiplicity_ks_grids_sigmai}.
Each panel represents a different set of model assumptions, specifically, 
a flared disk and Poisson distributed $\multiplicitydistribution$ (panel $a$)), 
a flat disk and Poisson distributed $\multiplicitydistribution$ (panel $b$)), 
a flared disk and exponential distributed $\multiplicitydistribution$ (panel $c$)), 
a flat disk and exponential distributed $\multiplicitydistribution$ (panel $d$)). 
The combined best-fit grid point is shown as a filled red circle for each set of simulations, along with the 1-sigma and 2-sigma probability contours. 
The only set of model systems which does not match the Q1$-$Q16 Kepler observations well, 
is when the number of planets per star $\multiplicitydistribution$ is Poisson distributed with a flared planetary disk.
Only this combination of model assumptions requires a dichotomous planetary system population, i.e. the Kepler dichotomy. 
}
\label{fig:probgrid_combined} 
\end{figure*}

\subsection{Combining independent tests}

The results from comparing the $\detectedplanets$ and $\transdurationratio$ distributions can be combined
in order to estimate the overall best-fit parameters, for a given set of model assumptions.
The $p$ values from each test are combined using Fisher's method into a single test statistic,
\begin{equation}
\label{eq:chisqr_combined}
\chi^2_\text{combined} \approx -2 \sum_{m=1}^M \ln p_m
\end{equation}
where $M$ is the number of tests combined and $p_m$ is the p-value of the $m$th test. 
The degrees of freedom is given by $2M$, where in this case $M=2$.
We use this combined statistic to produce a probability grid, $P_\text{combined}$, across the parameter space for each set of model assumptions.
$P_\text{combined}$ is derived from the likelihood $P_{ij} \propto \exp (-\chi_{\text{combined},ij}^2/2)$ and the requirement $\sum P_{ij} = 1$.

Figure \ref{fig:probgrid_combined} displays the probability grids for each set of model assumptions, along with the best-fit point and the $1\sigma$ and $2\sigma$ probability contours.
Panel $a)$ of Figure \ref{fig:probgrid_combined} combines the tests of Figure \ref{fig:multiplicity_ks_grids_sigmai},
and Panel $b)$ combines the tests of Figure \ref{fig:multiplicity_ks_grids_z0}.
A similar process is involved for panels $c)$ and $d)$,
where the number of planets per star $N_{pl}$ is drawn from an exponential distribution.
The intermediate figures for these two panels are not displayed for succinctness. 

\section{Results and Discussion}
\label{sec:results}

\subsection{Flared Disk Model}
Our result for a flared disk with a Poisson distributed number of planets per star, 
appears to be compatible with the majority of previous analyses.
We find a mean number of planets per star $\meanplanets =2.0^{+0.3}_{-0.2}$,
over our parameter space, [3 days $<P<200$ days] and [1 $R_\oplus<R_p<5\;R_\oplus$].
For a similar orbital period and planet radius parameter space, \cite{Gaidos2016} find $\meanplanets=2.2\pm0.3$ for Kepler M dwarfs.
\cite{Fang2012a} report $\meanplanets\sim$1.5 for $R_p>1.5\;R_\oplus$,
where the reduction in $\meanplanets$ likely comes from the exclusion of planets with radii between $1.0\;R_\oplus < R_p < 1.5\;R_\oplus$.

Similarly, we find the mode of the Rayleigh distributed mutual inclinations is given by $\sigma_i = 2.3^{+0.9}_{-0.4}$ degrees, 
consistent with the bulk of previous results with $\sigma_i\sim2^\circ$
\citep{Lissauer2011,Tremaine2012,Figueira2012,Fang2012a,Johansen2012,Fabrycky2014,Ballard2014,Gaidos2016}.

We are unable to achieve a good match to the Kepler Q1$-$Q16 detections for a flared disk model,
contrary to the reported result by \cite{Fang2012a},
where a flared disk model reproduced the $\detectedplanets$ distribution without the need for a Kepler dichotomy.
The discrepancy likely comes from the unique $N_{pl}$ distribution chosen by \cite{Fang2012a}, a ``bounded uniform'' distribution.
The bounded uniform distribution is produced by first choosing a maximum number  of planets $n_\text{i,max}$ from a Poisson distribution, 
then choosing the number of planets in the system $n_i$ from a uniform distribution between 1 and $n_\text{i,max}$.

It has previously been shown that the Kepler $\detectedplanets$ distribution can be matched without the need for a Kepler dichotomy,
when $N_{pl}$ is drawn from an exponential distribution \citep{Gaidos2016}.
We find that the Kepler sample is consistent with $\sigma_i = 2.4^{+0.9}_{-0.5}$ degrees and $\meanplanets = 1.6^{+0.3}_{-0.2}$
drawn from an exponential distribution.
Here we disagree with \cite{Gaidos2016}, who preferred near-coplanar mutual inclinations.
While we also achieve good fits for near-coplanar orbital planes, 
comparing $\transdurationratio$ distributions strongly rules out mutual inclinations $\lesssim 1.4^\circ$.
This illustrates the importance of modeling both $\detectedplanets$ and $\transdurationratio$ distributions,
where \cite{Gaidos2016} only modeled the $\detectedplanets$ distribution.

\subsection{Flat Disk Model}

For a flat planetary disk model
\citep{DullemondMonnier2010}, 
a good fit to the Kepler candidates can be achieved when the number of planets per star is drawn from both a Poisson or exponential distribution.
That is, independent of the $N_{pl}$ distribution chosen, the flat disk model removes the need for a Kepler dichotomy. 
When $N_{pl}$ is drawn from a Poisson distribution, we find $Z_0 = 1.6^{+0.6}_{-0.4}\;R_*$ and $\meanplanets = 2.4^{+0.6}_{-0.4}$.
Notably, the mean number of planets per star $\meanplanets$ is consistent between the assumed planetary disk models. 

We use a flat planetary disk model with a Gaussian disk thickness $Z_0$. 
We can compare this value to the inner Solar System (Figure \ref{fig:solsys_heightaboveplane}).
For the inner Solar System planets, $Z_\text{max}\approx8\;R_*$, giving $Z_0\approx5\;R_*$, where $Z_0\approx2\, Z_\text{max}/\pi$. 
This is significantly larger than our derived value of $Z_0=1.6^{+0.6}_{-0.4}\;R_*$ for our sample of closely packed Kepler systems. 
This may give some indication of the flat disk model's applicability at larger semi-major axes,
or may be reflective of the different parameter spaces probed.

\section{Summary and Conclusion}
\label{sec:conclusion}
We estimate the inherent orbital period and planet radius distributions for the Kepler Q1$-$Q16 catalog,
within the parameter space [3 days $<P<200$ days] and $[1\;R_\oplus<R_p<5\;R_\oplus]$.
We find that both distributions are well described by broken power laws, with breaks occurring at $\sim$15 days and $\sim$$2.7\;R_\oplus$.
These inherent distributions are used to populate model planetary systems for flat and flared planetary disk models,
and for the number of planets per star $N_\text{pl}$ drawn from Poisson and exponential distributions.

We confirm that a flared planetary disk model with $N_\text{pl}$ drawn from a Poisson distribution, 
is not consistent with the Kepler detections. 
We also confirm that Kepler detections are well matched when $N_\text{pl}$ is drawn from an exponential distribution,
without the need to invoke a dichotomous planetary system population.
In this paper we use a flat inner planetary disk model, 
where planets with smaller periods tend to have larger inclinations.
When a flat rather than a flared planetary disk model is assumed,
model systems are consistent with Kepler detections, 
without the requirement of a Kepler dichotomy,
and independent of the chosen $N_\text{pl}$ distribution.

We find that the mean number of planets per star $\meanplanets$ is largely model independent,
$\sim$2.0 when $N_\text{pl}$ is drawn from a Poisson distribution,
and $\sim$1.6 when $N_\text{pl}$ is drawn from an exponential distribution,
for [3 days $<P<200$ days] and [$1\;R_\oplus<R_p<5\;R_\oplus$].
This contrasts with the Solar System where there are 0 planets within this parameter space.

Similarly, we find for a flared planetary disk model, mutual inclinations are distributed with a mode $\sim$$2.2^\circ$.
For a flat planetary disk model, the Gaussian disk thickness $Z_0\sim\,$1.5 $R_*$,
much lower than the $\sim$5 $R_*$ of the inner Solar System.

\subsection{The Kepler Dichotomy}

The underproduction of model systems with a single detected transiting planet has been well studied.
This has lead to the invocation of a dichotomous planetary system population,
where one population suppresses the number of detected transiting planets, 
resulting in a higher likelihood of producing a single detected transiting planet.
Many physical explanations for the existence of the dichotomy have been put forward \citep{Johansen2012,Weissbein2012,Hansen2013,Moriarty2015,Lai2016,Spalding2016}.

\cite{Dawson2016} generated sets of planetary systems with various gas depletion factors using N-body simulations of planetary embryos.
No set of simulations was a good match to the period ratio, $\Delta$ (Equation \ref{eq:delta}), planet multiplicity and $\xi$ distributions of the observed Kepler sample.
Some improvement was found when simulated planetary systems were allowed to be a mix of ``dynamically hot'' and ``dynamically cold systems''. 
However, this improvement becomes less pronounced when taking into account the partial correlations between these distributions, 
particularly between $\xi$ and $\Delta$.





It has also been shown that the requirement of the dichotomy is not robust to the assumed distribution for the number of planets per star \citep{Fang2012a,Gaidos2016}.
This is confirmed in this paper, and in addition, 
we show that a planetary system dichotomy is also not required for a flat inner planetary disk model.
This result is independent of the choice of distribution for the number of planets per star $N_p$.
We emphasize that we apply the flat planetary disk model only to the short period range of Kepler candidates.

Of the sets of model assumptions explored in this paper, the need for a Kepler dichotomy only exists for a flared inner planetary disk,
with the number of planets per star drawn from a Poisson distribution.

The Kepler dichotomy describes the apparent need for a dichotomous planetary system population,
with respect to a star's probability of producing multiple transiting planets.
We show that the Kepler dichotomy is only required under specific model assumptions.
Specifically, when the inner part of a planetary disk is assumed to be flared, 
while also requiring the number of planets per star to be Poisson distributed. 
When removing either or both of these assumptions, the need for a Kepler dichotomy disappears.

\bibliographystyle{mn2e} 
\bibliography{inclination_refs}

\appendix

\section{Inclination angles of planetary orbital planets}
\FloatBarrier
There are a number of different angles used in the literature which have all been referred to as the planet inclination. 
Where we have used an inclination angle, we have attempted to be as explicit as possible. 
The below figure illustrates different inclination angles used throughout the paper.

\begin{figure}
\includegraphics[width=0.45\textwidth]{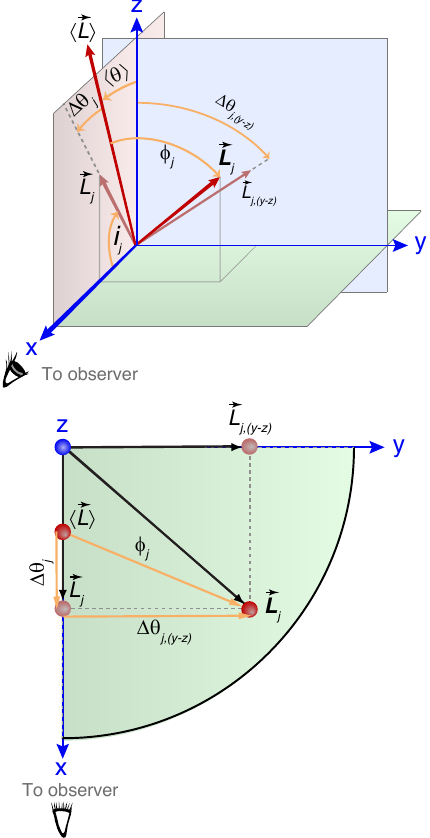} 
\caption{\emph{Top: }The inclination angles used in this paper, with the observer in the direction of the x axis.
We have chosen our coordinate system (without the loss of generality) such that the invariable plane vector $\vec{\langle L\rangle}$ is in the x-z plane.
The angular momentum vector of the planet is $\vec{\mathbf{L_j}}$ and the projection of $\vec{\mathbf{L_j}}$ onto the x-z plane is given by $\vec{L_j}$.
The $j$ subscript refers to an individual planet. 
The projected inclination of the $j^\text{th}$ planet's orbital plane relative to the observer is represented by $i_j$, typically reported by transit and radial velocity detections (e.g. $M\sin i_j$).
The coplanarity of planets refers to the distribution of $\phi_j$, the angle between the invariable plane of the system and the orbital planes of the planets.
The component of $\phi_j$ that the observer can probe is $\Delta\theta_j$ (whose distribution is plotted in the lower panels of Figure \ref{fig:model_illustration}), 
while $\Delta\theta_{j,(y\mhyphen z)}$ is the orthogonal component of the inclination which typically cannot be measured by the observer. \\
\emph{Bottom:} The coordinate system from the top panel viewed from above and compressed to two dimensions (the z axis points out of the page). 
The filled circles represent the tops of the vectors from the top panel.}
\label{fig:inclination_angles}
\end{figure}

\section{Planet distribution function}
\label{app:pldf}

Our planet distribution function (PLDF) has 7 free parameters, $F_0$, $\beta_1$, $\beta_2$, $P_\text{brk}$, $\alpha_1$, $\alpha_2$, $R_\text{brk}$.
\begin{align*}	
&\frac{df}{dP\,dR_p} = C\,F_0\, g(P,R_p) \numberthis\label{eq:model} \\ =
&\begin{cases}
C\,F_0\,P^{\beta_1} R_P^{\alpha_1} & \text{if }P < P_\text{brk}\text{ and }R_p < R_\text{brk}  \\ 
C\,F_0\,P^{\beta_1} R_P^{\alpha_2} R_\text{brk}^{\alpha_1-\alpha_2} & \text{if }P < P_\text{brk}\text{ and }R_p \ge R_\text{brk} \\
C\,F_0\,P^{\beta_2} P_\text{brk}^{\beta_1-\beta_2} R_P^{\alpha_1} & \text{if }P \ge P_\text{brk}\text{ and }R_p < R_\text{brk} \\
C\,F_0\,P^{\beta_2} P_\text{brk}^{\beta_1-\beta_2} R_P^{\alpha_2} R_\text{brk}^{\alpha_1-\alpha_2} & \text{if }P \ge P_\text{brk}\text{ and }R_p \ge R_\text{brk}
\end{cases}  
\end{align*}
where $F_0$ is the number of planets per star within our parameter space, $R_\text{brk}$ and $P_\text{brk}$ are the transition points between the two power laws for the planet radius and orbital period respectively. 
The normalization constant $C$ is calculated from the requirement
\begin{equation}
\int_{R_\text{min}}^{R_\text{max}} \int_{P_\text{min}}^{P_\text{max}} C\; g(P, R_p) \;dP \,dR_p = 1
\end{equation}
where the integration limits $R_\text{min}$, $R_\text{max}$, $P_\text{min}$ and $P_\text{max}$ are given in Equation \ref{eq:planet_sample}.

We follow \cite{Youdin2011} and \cite{Burke2015} by implementing a Poisson likelihood for our PLDF. By maximizing this likelihood we can obtain best-fit parameters for our model. 
\begin{equation}
\ln(L) \propto \left[ \sum^{N_\text{pl}}_{i=1}  \ln\left(C\, F_0\, g(P,R_P) \right)\right]- N_\text{exp}
\label{eq:loglikelihood}
\end{equation}
where $N_\text{exp}$ is the expected number of planet detections for the set of model parameters, and is given by
\begin{equation}
N_\text{exp} = C\, F_0 \int_{R_\text{min}}^{R_\text{max}} \int_{P_\text{min}}^{P_\text{max}}  \left[\sum_{j=1}^{N_*} \eta_j(P, R_p)\right] g(P,R_P)\; dP\;dR_p
\label{eq:nexp}
\end{equation}
where $\eta_j(P, R_p)=\eta_\text{detect}\times \eta_\text{transit}$ is the combined transit and pipeline detection efficiency of the $j$th star for the specified period and radius. 
The pipeline detection efficiency $\eta_\text{detect}$ is given by Equation \ref{eq:eta_disc} and the transit probability $\eta_\text{transit}=R_*/a$, where $a$ is the semi-major axis.

We calculate $\left[\sum_{j=1}^{N_*} \eta_j(P, R_p)\right]$ for a grid in orbital period and planet radii, in bins of 1.5 days and 0.05 $R_\oplus$ respectively.
For each grid point, we sum over all stars in our sample. 
The mean combined transit and pipeline detection efficiency $\langle\eta(P,R_p)\rangle$ can then be found be dividing this term by the number of stars in our sample, $N_*$.

\section{Simulated planetary systems}
\subsection{Testing the stability of sequential planet pairs}
\label{sec:stability}
The dynamical spacing $\Delta$ describes the separation of two planets in units of their mutual Hill radius. 
The mutual Hill radius of two planets is given by
\begin{equation}
R_{H,ij}=\left(\frac{m_i+m_j}{3M_*}\right)^{1/3}\frac{a_i+a_j}{2}
\end{equation}
where $m_i$ and $m_j$ are the planet masses for the inner and outer planets respectively. 
The dynamical spacing $\Delta$ is the semi-major axis spacing of the two planets, in units of the mutual Hill radius.
\begin{equation}
\Delta_{ij}=\frac{a_j-a_i}{R_{H,ij}}
\label{eq:delta}
\end{equation}
where $a_i$ and $a_j$ are the semi-major axes of the inner and outer planets respectively.
Analytic stability solutions exist for a system which contains exactly two planets, $\Delta_{ij}\gtrsim3.46$ \citep{Gladman1993,Chambers1996}, 
although it is not possible to ensure this requirement for our simulated systems\footnote{
Although our simulated systems may produce exactly two planets within our parameter space, 
we cannot rule out the possibility of additional planets outside of our parameter space, 
which would invalidate the analytic solution.}.
For systems with $\numinsys \ge 3$, we use an empirical stability criteria for two adjacent planet pairs (three sequential planets). 
A set of three sequential planets with indicies $i$, $j$, and $k$ is deemed unstable when
\begin{equation}
\Delta_{ij}+\Delta_{jk} < 18
\end{equation}
where $\Delta_{ij}$ and $\Delta_{jk}$ are the dynamical spacing of the inner and outer planet pair from the three sequential planets \citep{Lissauer2011}.
If there are only two simulated planets in a system, $\Delta_{ij}<10$ results in the system being labeled unstable. 

Should any set of planets fail the above stability criteria, the system is deemed unstable and new planet periods are redrawn for all $\numinsys$ planets as in step 5. 
New planetary radii are not redrawn, since passing the stability criteria is biased towards sets of planets with small planetary radii, 
where stability is more easily achieved. 
Redrawing planetary radii immediately would result in a simulated $R_p$ distribution skewed towards small $R_p$, relative to the underlying distribution in Section \ref{sec:period_radius_distributions}. 
Should the stability criteria fail $10^3$ times for the same set of planetary radii, new $R_p$ and periods for all $\numinsys$ planets are redrawn as in step 4.

\subsection{Orbital plane inclinations relative to the observer}
\label{sec:assign_inclinations}
Once stability has been established for a model system, each planet is then assigned an inclination relative to the observer, 
the $i$ variable commonly seen in transit and radial velocity detections. 

For Rayleigh distributed mutual inclinations (flared disk in Figure \ref{fig:model_illustration}), $i$ is assigned as follows. 
An inclination $\phi$ is drawn from a Rayleigh distribution with mode $\sigma_\phi /\sqrt{2}$, where $\sigma_\phi$ is the mode of the Rayleigh distributed mutual inclinations. 
The factor of $1/\sqrt{2}$ is a conversion factor between the Rayleigh distributed mutual inclinations, 
and the Rayleigh distributed planet inclinations around the invariable plane.
The orbital plane of the planet is then rotated by a random uniform angle $\Omega$, giving
\begin{equation}
i_\text{flare}=\invplane+\phi\,\cos(\Omega).
\end{equation}

For a flat disk (Figure \ref{fig:model_illustration}), the perpendicular height above the invariable plane $Z_0$ is drawn from a Gaussian distribution with a mean of 0 and standard deviation $\sigma_Z$, in units of stellar radii. 
For a flat disk, unlike a flared disk, the assigned inclination $i$ is dependent on the semi-major axis of the planet. 
Again, the orbital plane of the planet is rotated by a random uniform angle $\Omega$, to account for a random viewing angle.
\begin{equation}
i_\text{flat}= \invplane + \arcsin(Z_0/a)\,\cos(\Omega)
\end{equation}
resulting in a tendency for larger inclinations for close-in planets and vice-versa (right planel of Figure \ref{fig:model_illustration}).

\section{Orbit-normalised transit duration ratio $\transdurationratio$}
\label{app:xi}
For a planet which transits through the centre of its star:
\begin{equation}
2R_*\approx v_\text{orb}T_\text{dur}
\end{equation}
where $v_\text{orb}$ and $T_\text{dur}$ represent the orbital velocity (assuming a circular orbit) and the transit duration of the planet respectively.
Note that for the Kepler sample, the simplification of a circular orbit is justified since $\xi$ is weakly dependent on eccentricity \citep{Fabrycky2014}.
In addition, eccentricity values for the Kepler sample are generally found to be associated with near-circular orbits (e.g. \cite{Fabrycky2014,Hadden2014,Dawson2016}), or with mean values around $\sim$0.1 \citep{Moorhead2011,Hansen2013}.

When the transit is not through the centre of the star $\left(2R_* = 2\sqrt{R_*^2-b^2}\right)$:
\begin{equation}
2\sqrt{R_*^2-b^2} = v_\text{orb}T_\text{dur}
\end{equation}
where $b$ is the impact parameter representing the transiting planet.
For a pair of planets which transit the same host star:
\begin{align}
\label{eq:xi_planet1}
&2\sqrt{R_*^2-b_\text{in}^2}=T_\text{dur,in} v_\text{orb,in}\\
\label{eq:xi_planet2}
&2\sqrt{R_*^2-b_\text{out}^2}=T_\text{dur,out} v_\text{orb,out}
\end{align}
where the "in" and "out" subscripts represent the inner and outer planets respectively. From Kepler's 3rd law:
\begin{align*}
&v_\text{orb,in}\propto \frac{a_\text{in}}{P_\text{in}}\propto P_\text{in}^{-1/3}
\end{align*}
Dividing \ref{eq:xi_planet1} by \ref{eq:xi_planet2}:
\begin{equation}
\label{eq:xi_cosratio}
\frac{\sqrt{R_*^2-b_\text{in}^2}}{\sqrt{R_*^2-b_\text{out}^2}} = \frac{T_\text{dur,in}/P_\text{in}^{1/3}}{T_\text{dur,out}/P_\text{out}^{1/3}}
\end{equation}
The RHS ratio is particularly useful for planetary transits as it is composed of well-measured variables. Setting the RHS to $\xi$ \citep{Steffen2010}: 
\begin{equation}
\label{eq:xi}
\xi=\frac{T_\text{dur,in}/P_\text{in}^{1/3}}{T_\text{out,in}/P_\text{out}^{1/3}}
\end{equation}
From \ref{eq:xi_cosratio}, a coplanar planetary pair will only give $\xi=1$ if the invariable plane (Fig \ref{fig:inclination_angles}) is exactly edge-on to the observer. For inclined invariable planes, a coplanar planetary pair will give $\xi>1$, as $b_\text{out} > b_\text{in}$. Values of $\xi<1$ are due to $b_\text{out}<b_\text{in}$, and are not possible in cases of perfect coplanarity. 

\begin{landscape}
\section{Previous coplanarity studies}
\label{sec:prev_studies}
\begin{threeparttable}
	\small
	\caption{\normalsize Comparison of exoplanet coplanarity studies}
	\label{tab:prev_studies}
	\setlength{\tabcolsep}{4pt}
	\begin{tabular}{cccccccccccc}
		\midrule
		& & & & & \multicolumn{2}{|c|}{Planet sample} \\
		\cmidrule(lr){6-7}
		Reference & $\Delta\phi$ distribution & Observables & Dispersion$^a$ & Sample (quarter, multiplicity) & Period (days) & Radius ($R_\oplus$) & Stellar sample & Dichotomy$^g$\\
		\midrule
		\cite{Lissauer2011} & Rayleigh & $\detectedplanets^{\,b}$ & $\sigma_\phi\sim2.0^\circ$ & Kepler (Q2, 1-6) & $3-125$ & $1.5-6$ & FGK dwarfs & 2.8\\
		\cite{Tremaine2012} & Fisher & $\detectedplanets$ & $\sigma_\phi^c <4.0^\circ$ & RV \& Kepler (Q2, 1-6) & $<200$ & $< 22$ & FGK dwarfs & - \\
		\cite{Figueira2012} & Rayleigh & $\detectedplanets$ & $\sigma_\phi^d \sim 1.4^\circ$ & HARPS \& Kepler (Q2, 1-3) & $<50$ & $>2$ & FGK dwarfs & -  \\
		\cite{Fang2012a} & Rayleigh, R of R & $\detectedplanets$, $\transdurationratio^{e}$ & $\sigma_\phi^c\sim1.4^\circ$ & Kepler (Q6, 1-6) & $<200$ & $1.5-30$ & FGK dwarfs & 1 \\
		\cite{Johansen2012} & uniform $i$ + rotation$^{\,f}$ & $\detectedplanets$ & $\sigma_\phi<3.5^\circ$ & Kepler (Q6, 1-3)  & $<240$ & $<22$ & FGK dwarfs & 3\\
		\cite{Weissbein2012} & Rayleigh & $\detectedplanets$ & no fit & Kepler (Q6, 1-6)  & $<75^h$ & $-$ & FGK dwarfs & -\\
		\cite{Hansen2013} & Rayleigh & $\detectedplanets$ & - & Kepler (Q6, 1-6) & $<1.1$ AU & - & - & 2\\
		\cite{Fabrycky2014} & Rayleigh & $\transdurationratio$ & $\sigma_\phi\sim1.8^\circ$ & Kepler (Q6, 1-6) & $<130^h$ & $-$ & FGK dwarfs & - \\
		\cite{Ballard2014} & Rayleigh & $\detectedplanets$ & $\sigma_\phi = 2.0^{\circ\,+4.0}_{\;\;\;-2.0}$ & Kepler M-dwarfs (Q16, 1-5) & $1-200$ & $-$ & M stars & 3\\
		\cite{Gaidos2016} & Rayleigh & $\detectedplanets$ & $\sigma_\phi \sim 0^\circ$ & Kepler M-dwarfs (Q16, 1-5) & $<180$ & $1-4$ & M stars & - \\
		This paper & Rayleigh / Flat disk & $\detectedplanets, \transdurationratio$ & $1.6^{+0.6}_{-0.3}$ & Kepler (Q16, 1-6) & $3-200$ & $1-5$ & FGK dwarfs & - \\
		\hline
	\end{tabular}
	\begin{tablenotes}[]
		\item[a] The mode of the Rayleigh distribution of $\phi$ values (Fig. \ref{fig:inclination_angles}) around the invariable plane.
		\item[b] $\detectedplanets$ is the multiplicity vector for the numbers of observed k-planet systems, 
		i.e. $\detectedplanets = [\hat{N_1}, \hat{N_2}, \hat{N_3},...]$.
		\item[c] Converted from the mean $\mu$ of the mutual inclination Rayleigh distribution: $\sigma_\phi=\sqrt{2/\pi} \:\: \sigma_i$.
		\item[d] Converted from Rayleigh distribution relative to the invariable plane:  $\sigma_\phi=\sqrt{2} \;\sigma_{\Delta\theta}$.
		\item[e] $\transdurationratio$ is the normalized transit duration ratio (Appendix \ref{app:xi}) as given in \cite{Steffen2010}.
		\item[f] Each planet is given a random uniform inclination between $0^\circ-5^\circ$. This orbital plane is then rotated uniformally between $0-2\pi$ to 
		give a random longitude of ascending node.
		\item[g] The factor by which the number of simulated 1-planet systems are lower than observed
		\item[h] Converted from a maximum semi-major axis, assuming a Solar mass star
         \end{tablenotes}
\end{threeparttable}
\label{lastpage}
\end{landscape}

\end{document}